\documentclass[pageno]{jpaper}

\usepackage{soul, xcolor}
\setstcolor{red}
\usepackage{booktabs} 
\usepackage{graphicx}
\usepackage{color}
\usepackage[all]{nowidow}

\usepackage{color,verbatim,afterpage,url,xfrac, algorithm}
\usepackage{algpseudocode}
\graphicspath{{figures/}}

\long\def\comment#1{}

\def\parah#1{\vspace*{0.0in} \noindent{\bf #1:}}
\newcommand{\red}[1]{ \textcolor{red}{#1}}

\newcommand{\qnl}{\textsc{QNL8QR-v5\ }}
\newcommand{\cn}{\textsc{CNOT\ }}

\usepackage{graphicx}
\usepackage{subscript}
\usepackage{MnSymbol,wasysym}
\usepackage{listings}
\usepackage{multirow,rotating}
\usepackage{fancyhdr} 
\usepackage{moresize}
\usepackage{alltt}
\usepackage{tikz}

\date{}

\graphicspath{{figures/}}

\usepackage[normalem]{ulem}

\usepackage{booktabs}   

\usepackage{braket}
\usepackage{physics}

\usepackage{listings}

\newcommand{\basetriangle}{
  \def\h{0.866025}
  \coordinate (X) at (-1/2,{-1/3*\h});
  \coordinate (Y) at (1/2,{-1/3*\h});
  \coordinate (Z) at (0,{2/3*\h});
  \fill[white!10, draw=black] (X) -- (Y) -- (Z) -- cycle;
  \filldraw (X) circle (3pt); \filldraw (Y) circle (3pt); \filldraw (Z) circle (3pt);
  }

\begin{document}

\title{Heuristics for Quantum Compiling with a Continuous Gate Set}        
\author{Marc Grau Davis, Ethan Smith, Ana Tudor, Koushik Sen, Irfan Siddiqi, Costin Iancu}
\date{} 
\maketitle
\thispagestyle{empty}

\begin{abstract}
 {\it We present an algorithm for compiling arbitrary unitaries into a
 sequence of  gates native to a quantum processor. As accurate \cn gates are
 hard for the foreseeable
 Noisy-Intermediate-Scale Quantum devices era, our  A* inspired algorithm
 attempts  to minimize their count, while accounting for connectivity. We discuss the search strategy
 together with ``metrics'' to expand the solution frontier.  For a workload
of circuits with complexity appropriate for the NISQ era, we produce solutions well within the best upper
 bounds published in literature and match or exceed hand tuned
 implementations, as well as other existing synthesis alternatives. In
 particular, when comparing against state-of-the-art available
 synthesis packages we show $2.4\times$ average (up to $5.3\times$) reduction in CNOT count. We also show how to re-target the algorithm for a different
 chip topology and native gate set, while obtaining similar quality
 results. We believe that empirical tools like ours can
 facilitate algorithmic exploration, gate set discovery for quantum processor designers,
as well as providing useful optimization blocks within the quantum compilation tool-chain.}
\end{abstract}

\section{Introduction}
\label{sec:intro}

There is a high probability that quantum computing will deliver
transformational scientific results within the next few decades. Right
now, we are in an era of effervescence, where the first available~\cite{IBM-2017,Google-2018,Intel-2018}
hardware implementations of quantum processors have opened the doors
for exploration in quantum hardware, software and algorithmic
design. \comment{At the hardware level, chip designers keep introducing new
primitives  motivated by the desire  to model physical phenomena,
such as qutrits~\cite{qtrit,bhscramble} used 
to model information scrambling
in black holes. As current generation chips have relatively short
decoherence times, compile time  optimization of quantum circuits for depth
reduction is paramount to success in the near future. At the
algorithmic level, there exists a dire need for discovery combined
with a need for tools to assist in mapping a science level 
Hamiltonian to the equivalent circuit.}
All three lines of inquiry  have in common that obtaining the
unitary matrix associated with the transformation (algorithm, gate,
circuit etc.) is ``easy'', while deriving equivalent circuits from said
unitary is hard.  Quantum circuit synthesis is an approach to derive a good circuit
that implements a given unitary and  can thus facilitate advances
in all these directions: hardware, software and  algorithmic exploration. 

Research into quantum circuit synthesis has a long~\cite{DawsonNielson05,Nagy16,ola15, ZXZ16,MIM13,Qcompile16,kmm13,GS13,KAK} history.
We believe that synthesis can be a tool of great utility in the
quantum development kit for the Noisy Intermediate-Scale Quantum (NISQ)
Devices era, which is characterized by design space exploration at
small qubit scale, together with a need for highly optimized
implementations of circuits.  To foster adoption,
synthesis tools
need to overcome some of the currently perceived
shortcomings:
\begin{itemize}
\item Synthesized circuits tend to be deep
\item Synthesis does not account for hardware
  topology 
\item The compilation itself is slow
\end{itemize}

In this paper we describe a  pragmatic heuristic synthesis
algorithm, whose goal is to minimize the number of \cn gates used in the resulting
circuit. As \cn has low fidelity on existing hardware and it is
expected to be the limiting factor in the near future of NISQ devices, this metric has been targeted by
others~\cite{raban,qaqc,synthcsd,KAK}.

The algorithm is inspired by the A*~\cite{astar} search strategy and works as
follows. Given the unitary associated with a quantum transformation,
we attempt to alternate layers of single qubit gates and \cn
gates. For each layer of single qubit gates we assign the
parameterized single qubit unitary to all the qubits. We then try to place a \cn
gate wherever the {\it chip connectivity allows}, and add another layer of single qubit gates. We pass the parameterized
circuit into an optimizer~\cite{Powell1994}, which instantiates the parameters
for the partial solution such that it minimizes a distance
function. At each step of the search, the solution with the shortest
``heuristic''  distance from the original unitary is expanded. The algorithm stops when the current
solution is within a small threshold distance from original.  We now
have a concrete description of a circuit that can be implemented on
hardware. 

We target two superconducting qubit architectures: the \qnl chip developed by the UC Berkeley
Quantum Nanoelectronics Laboratory ~\cite{qnl}, with eight
superconducting qubits connected in a line topology and the IBM
Q5~\cite{ibmq5}  chip with qubits connected in a ``bowtie''. Both
chips have a similar native gate set composed of single qubit
rotations and CNOT gates. For evaluation we use  known algorithms and gates published in
literature, e.g. QFT, HHL, Fredkin, Toffoli etc., with implementations
obtained from other researchers~\cite{noisemap}.

Overall, we believe that we have made several good contributions that
advance the state of the art in quantum circuit synthesis. The results indicate that synthesis can be a very useful tool in the
stack of quantum circuit compilation tools. When comparing against circuits that were painstakingly
hand optimized, our implementation matches and sometimes reduces the
\cn count.
When comparing against state-of-the-art available tools such as
UniversalQ~\cite{uq}, our implementation produces much better
circuits, with $2.4 \times$ average reduction in CNOT gates, and by as much as $5.3
  \times$. The data dispels the concern that synthesis produces
  deep circuits.   

To our knowledge we provide  the first practical demonstration of good topology aware
synthesis. Intuitively, by specializing the search strategy for a
given topology results in circuits than may not need additional SWAP
operations inserted at the mapping stage.  Existing approaches assume
all-to-all connectivity, and modifications  to handle restricted
topologies introduce large (e.g. $4\times$~\cite{raban})
proportionality constants. In our case we observe only modest
differences between circuits synthesized for all-to-all (bowtie) and
circuits synthesized on a linear topology. We observe
reduced \cn count on five circuits (half workload), with an average of 15\% reduction
for the whole workload. Furthermore, 
the depth  difference  from topology customization cannot be recuperated by the rest of the
optimization toolchain: the final depth  of a circuit synthesized
  for the fully connected topology and optimized and mapped for the linear topology by IBM
  QISKit, is longer than the depth of the circuit synthesized directly
  for the linear topology. We observe a 53\% average increase in
  depth, and up to $4\times$.

We also show how our infrastructure can be easily retargeted to
different native gate sets and qutrit~\cite{qtrit}
based circuits.  To our knowledge this is the first demonstration of
synthesis of multi-gate  multi-qutrit based circuits.

The rest of this paper is structured as follows. In
Section~\ref{sec:bg} we introduce the problem, its motivation and
provide a short primer on quantum computing. In
Section~\ref{sec:alg} we describe our algorithm and its
implementation, while in Section~\ref{sec:use} we present results for
the three usage scenarios. In Section~\ref{sec:related} we describe the related work,
while in Section~\ref{sec:disc} we discuss future uses of synthesis in
the NISQ era. 
\section{Background }
\label{sec:bg}

In quantum computing, a qubit is the basic unit of quantum information. Physically, qubits are two-level quantum-mechanical systems, whose general
quantum state is represented by a linear combination of two orthonormal basis states (basis vectors).  The most common basis is the equivalent
of the 0 and 1 values used for bits in classical information theory, \\
respectively $\ket{0} = \begin{bmatrix} 1 \\ 0 \end{bmatrix}$ and
$\ket{1}$ = $\begin{bmatrix} 0 \\ 1 \end{bmatrix}$. The generic qubit state is a superposition of the basis states, i.e. $\ket{\psi} = \alpha \ket{0} + \beta \ket{1}$, with $\alpha$ and $\beta$ complex amplitudes such as $|\alpha|^2+|\beta|^2=1$.

\comment{Although it appears there are four degrees of freedom in the generic qubit state $\ket{\psi}$, the normalization constraint removes one degree of freedom. Thus qubits can be represented in Hopf coordinates~\cite{Hopf1931} with three degrees of freedom $\alpha = e^{i\psi}cos\frac{\theta}{2}$ and $\beta = e^{i(\psi+\phi)}sin\frac{\theta}{2}$. Furthermore, since the phase of the $e^{i\psi}$ has no physical observability, the notation can be further simplified to two degrees of freedom, with $\phi$ the significant relative phase: $\alpha = cos\frac{\theta}{2}$ and $\beta = e^{i\phi}sin\frac{\theta}{2}$. This is usually visualized using the Bloch sphere.}

\subsubsection{Gate Sets in Quantum Computing} The prevalent model of quantum computation is the circuit model introduced by~\cite{qcircuit}, where information carried by qubits  (wires) is modified by quantum gates, which mathematically correspond to unitary operations. A complex square matrix U is {\bf unitary} if its conjugate transpose $U^*$ is also its inverse, i.e. $UU^* = U^*U = I$.

In the circuit model, a single qubit gate is represented by a $2 \times 2$ unitary matrix U. The effect of the gate on the qubit state is obtained by
multiplying the U matrix with the vector representing the quantum state $\ket{\psi'} = U\ket{\psi}$.

The most general form of the unitary associated with a single qubit
gate is  the ``continuous''
or  ``variational'' gate representation. \begin{equation} \label{eq:u3} U3(\theta,\phi,\lambda) = \begin{pmatrix} cos{\frac{\theta}{2}}  &
  -e^{i\lambda}sin{\frac{\theta}{2}} \\
  e^{i\phi}sin{\frac{\theta}{2}} &
  e^{i\lambda+i\phi}cos{\frac{\theta}{2}}\end{pmatrix}\end{equation}

In quantum computing theory, a set of quantum gates is {\it universal} if any computation (unitary transformation) can be approximated on any number of qubits to any precision when using only gates from the set. On the hardware side, quantum processors expose a set of native gates which constitute an universal set.  Quantum processors built from superconducting qubits usually provide a gate set consisting of single qubit rotations ($R_x$, $R_y$, and $R_z$) and two qubit \cn gates. \comment{  The unitaries for single qubit rotation gates are
\begin{equation} \label{eq:rs}
R_x(\theta) =  \begin{pmatrix} cos{\frac{\theta}{2}}  & -i sin{\frac{\theta}{2}} \\  -i sin{\frac{\theta}{2}} & cos{\frac{\theta}{2}} \end{pmatrix}, R_y(\theta)= \begin{pmatrix} cos{\frac{\theta}{2}}  & -sin{\frac{\theta}{2}} \\   sin{\frac{\theta}{2}} & cos{\frac{\theta}{2}} \end{pmatrix},  R_z(\phi)= \begin{pmatrix} e^{-i{\frac{\phi}{2}}}  & 0\\   0 &  e^{i{\frac{\phi}{2}}}\end{pmatrix}
\end{equation}
}

A \cn, or controlled NOT gate, flips the target qubit {\it iff} the control qubit is $\ket{1}$ and it has the following unitary \begin{equation} \label{eq:cn} CNOT = \begin{pmatrix}  1 & 0 & 0 & 0 \\ 0 & 1 & 0 & 0 \\ 0 & 0 &0 & 1 \\ 0 & 0 & 1 & 0 \end{pmatrix}\end{equation}

A circuit is described by an  evolution in space (application on qubits) and time of gates.   Figure~\ref{fig:circ} shows an example circuit that applies single qubit and \cn gates on three qubits.

\subsection{Background on Quantum Circuit Synthesis}

\begin{figure}
 \comment{ [width=3in,height=1in]}
 \centerline{\includegraphics[width=2in,height=.6in,keepaspectratio]{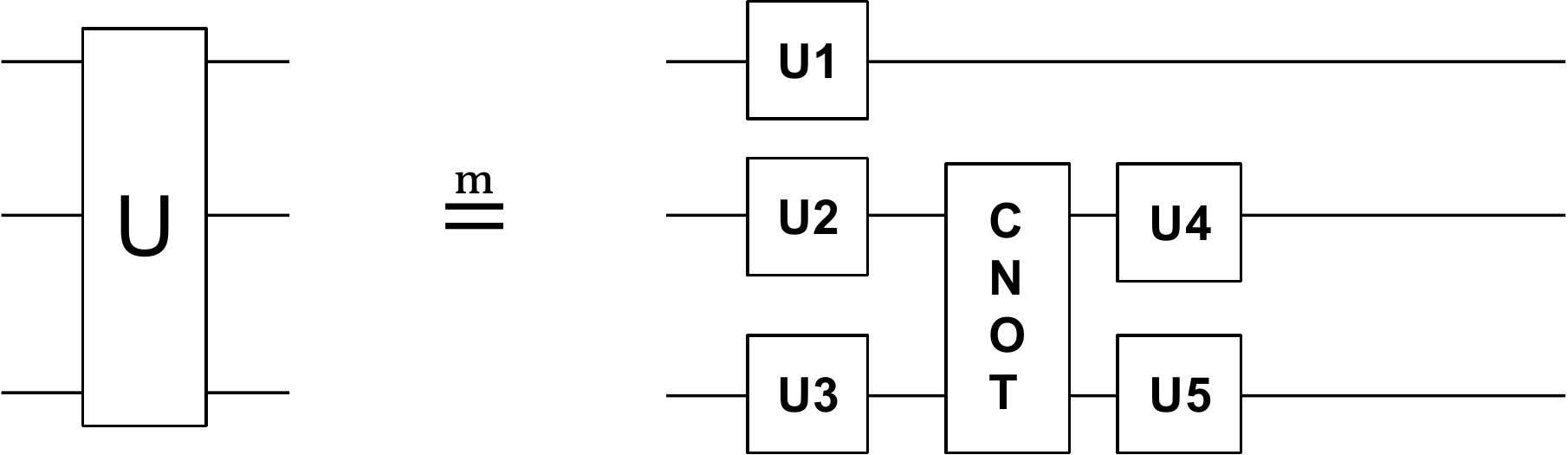}}
\centerline{ \includegraphics[width=2in,height=.4in,keepaspectratio]{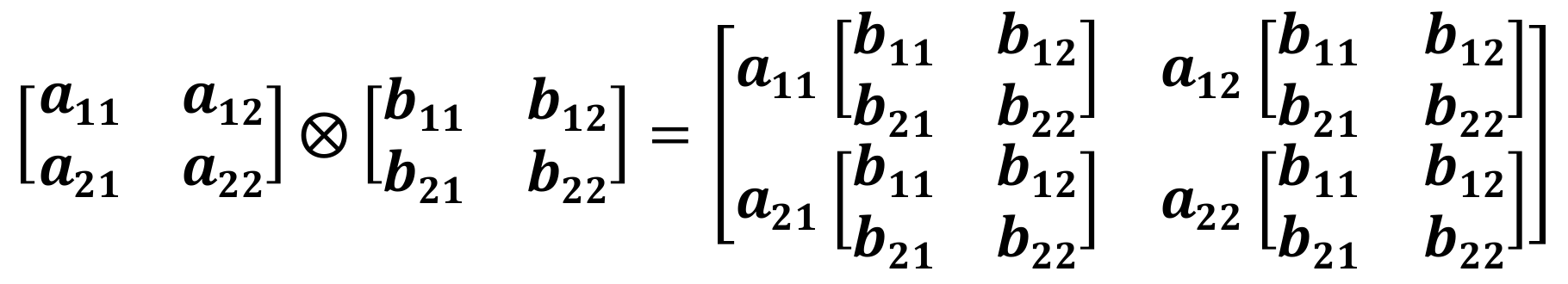}}
  \caption{\label{fig:circ} \it \footnotesize Unitaries (above) and tensors products (below). The unitary U represents a $n=3$ qubit transformation, where  U is a $2^n \times 2^n$  ($8 \times 8$) matrix. The unitary is
    implemented (equivalent or approximated) by the circuit on the right hand side. The single qubit unitaries are $2 \times 2$ matrices, while CNOT is a
    $2^2 \times 2^2$ matrix. The computation performed by the circuit is $(U3 \otimes U4 \otimes U5)(I2 \otimes CNOT)(U1 \otimes U2 \otimes U3)$, where   I2 is the identity $2 \times 2$ matrix and $ \otimes $ is the tensor product operator. The right hand side shows the tensor product of $ 2 \times 2$ matrices.}
 \end{figure}

A quantum transformation (algorithm, circuit)  on $n$ qubits is represented by a unitary matrix U of size $2^n \times 2^n$. The goal of circuit synthesis is to decompose U into a product of terms, where each individual term captures the application of a quantum gate on individual qubits.  This is depicted in Figure~\ref{fig:circ}.
The quality of a synthesis algorithm is evaluated by the circuit depth it produces (number of terms) and by the distinguishability of the solution from the original unitary.
We discuss in more detail related work in synthesis in Section~\ref{sec:related} and summarize in this section only the pertinent state-of-the-art results for  NISQ devices.

Circuit depth provides the optimality criteria for synthesis algorithms: shorter depth is better. \cn count is a direct indicator of overall circuit length, as the number of single qubit generic gates  introduced in the circuit is proportional with a constant given by decomposition rules. Thus \cn count or circuit depth can be used interchangeably when discussing optimality criteria. As \cn gates are problematic on NISQ devices, state-of-the-art approaches~\cite{raban,synthcsd} directly attempt to minimize their count. When reasoning about single qubit unitary operations, the $ZYZ$ decomposition rule states that any unitary $U$ can be rewritten as $U=e^{i\alpha}R_x(\beta)R_y(\gamma)R_z(\delta)$, with a proof available in~\cite{Nielsen-2010}; thus synthesis can focus on minimizing \cn count.

\comment{Reasoning whether a synthesis algorithm is exact is subtle in particular on NISQ devices. For example,~\cite{ctmq} describe an algorithm for exact synthesis using Clifford + T gates. On current devices, T gates are not native and implemented as approximations, thus their algorithm ultimately may produce an approximation. \comment{In our work we target only gate sets native to a particular quantum processor implementation. }\red{could mention that error due to imperfect hardware of real devices is way bigger than the theoretical difference between "exact" and good approximate circuits.  I could gather data to back this up, such as running "exact" vs our approximate QFT on IBM on their theoretical setup vs actual hardware.}}

There are two types of synthesis approaches: unitary decomposition using linear algebra techniques or empirical search based techniques. The state-of-the-art linear algebra techniques use Cosine-Sine Decomposition~\cite{raban,synthcsd} and provide  upper bounds on circuit depth. We use the tightest published upper bounds for the evaluation of our approach, as well as direct comparisons with the UniversalQ~\cite{uq} compiler, which implements these algorithms. Empirical approaches use search heuristics for decomposition. From these, we are mostly interested in numerical optimization approaches~\cite{ionsynth,qaqc} which are similar in spirit to our proposed solution. These tend to generate shorter circuits, but implicitly assume full qubit connectivity. We do not have access to these implementations, thus we can provide only indirect comparisons.  As stated, existing  algorithms are not widely used due to generating long circuits and not being able to take chip topology into account. The only exception is the ubiquitous deployment of KAK~\cite{KAK}  decompositions in commercial~\cite{qiskit,cirq,pyquil} compilers: KAK provides optimal decomposition of two qubit unitaries.

{\it Table~\ref{tab:depth} presents the best known upper bounds on \cn count for synthesis algorithms.} Note that for three qubits the bound is 20 CNOT, while for four qubits it is 100 CNOT.  Asymptotically, the tightest bound is introduced by~\cite{raban} to a \cn count of $0.16*(4^m+2*4^n)$. Because of the exponentiation, for current generation devices it is important to demonstrate quantitatively that we can attain shorter depth.

{\it Taking chip qubit connectivity into account} during synthesis affects circuit depth.  Most algorithms implicitly assume full qubit connectivity. Topology agnostic approaches may place \cn gates between qubits that are not physically connected. In these cases, the back-end compilers need to introduce $SWAP$ gates, each $SWAP$ gate being implemented using three \cn gates. Recent approaches try to provide bounds when specializing for topology, by estimating the number of additional SWAPs. The algorithms presented by~\cite{synthcsd} increase the \cn count by a factor of nine when restricting topology to a nearest-neighbor (linear topology) interaction, while~\cite{raban} claim a factor of four.  

\comment{Note that these are bounds derived when restricting topology from all-to-all to nearest-neighbor, and their generalization\footnote{Intuitively, the bound is likely to be lower for higher connectivity.}  to other topologies such as the IBM Q bow-tie is not clear. Furthermore, even when these bounds for topology specialization are available, not enough quantitative data is available to judge the quality of a synthesis algorithm. The topology agnostic bounds seem to be loose and most, if not all, studies do not present data when compiling realistic circuits or algorithms. The little data that is available  is provided for random unitaries.

When specializing for topology the bounds are even looser. For a $m$ qubit unitary, existing algorithms can generate circuits with ancilla qubits, which are extra qubits that are not used in the input or output of the program, but may be used in intermediate computation. Compiling with ancilla qubits results in $n > m$ qubit circuits.  Intuitively these should have a shorter length than the non-ancilla synthesis when topology is restricted, regardless of their theoretical higher bound.
We did not find much quantitative data for sample circuits in these cases, and there seem to be little guidelines for approaching the problem.  \red{we should probably do a little more research on compiling with ancillas if we are going to mention it}}

 \begin{table}[h]
\centering
{\small
\begin{tabular}{l|l|l|l|l|l}
n \ m       & 0 & 1& 2&3&4 \\ \hline
2      & 1& 2&{\bf 3}&-&- \\
3      & 3 &9& 14 &{\bf 20}& -  \\
  4     & 8 &22 &54&73 &{\bf 100}  \\

\end{tabular}
}
\caption{\label{tab:depth} \it Upper bound on CNOT gate count when synthesizing a $m$ qubit circuit into  $n$ qubits, with $m \le n$. Data is presented by ~\cite{raban}. The counts for $n=m$ are  introduced by ~\cite{synthcsd}. The counts for state preparation ($m = 0$) on two and three qubits are presented by ~\cite{znidaric}, and the count for state preparation on four qubits is introduced by~\cite{raban}. The generalization and upper bound of  is derived by~\cite{raban}. Note that the CNOT counts grow very fast. For example, the upper bound on any unitary on 10 qubits is about about 500,000 CNOT gates.}

\vspace{-.2in}
\end{table}

\subsubsection{Reasoning About Circuits and Algorithm Equivalence}

A quantum transformation can be implemented by multiple distinct quantum circuits.  that is when reasoning in terms of unitaries, there exist multiple decompositions of the unitary into terms that represent gates. Furthermore, when running on hardware, the unitary executed is often subtly different from the intended unitary.

Thus, it is often the case where we want to perform a particular quantum operation $A$ and because of external constraints we end up performing an approximation $B$, where $B \ne A$. Deciding which algorithm has executed is often referred to as distinguishability and  several metrics with operational motivation have been proposed.  Trace distance and fidelity~\cite{gilchrist,breuer,Nielsen-2010} have been proposed for distinguishing states. Metrics such as the diamond norm~\cite{Kitaev:2002} have been introduced to distinguish processes (algorithms).

Synthesis algorithms use norms  to assess the solution quality, and their goal is to minimize $\norm{U - U_S}$, where $U$ is the unitary that describes the transformation and $U_S$ is the computed solution. They choose an error threshold $\epsilon$ and use it for convergence, $\norm{U-U_S} \le \epsilon$. Early synthesis algorithms use the diamond norm, while more recent efforts~\cite{qaqc,HSnormsynth} use the Hilbert-Schmidt  inner product between the conjugate transpose of $U$ and $U_s$. This is motivated by its lower computational overhead. 
\begin{equation} \label{eq:hsn}
   \langle U, U_s \rangle_{HS} = \Trace(U^{\dag} U_s)
\end{equation}

\subsection{Quantum Processors}
\label{sec:qps}

Depending on the qubit technology, quantum processors may support different native gate sets, and  qubits may connected in different topologies. We target
processors with superconducting qubits since they implement a variety of topologies~\cite{qnl,ibmq5,tqfrigetti,Google-2018}  and are easier available. Most offer  a native
gate set consisting of rotations and \cn gates \{$R_x(90),
  R_z(\theta), CNOT$\}. We believe that our results are easily generalized across superconducting qubit architectures which tend to support rotations and a single two qubit gate (\cn, CRZ or SWAP).

\comment{We demonstrate synthesis results on two distinct architectures built using transmon~\cite{transmon} superconducting qubits. We initially  targeted the \qnl chip developed by the UC Berkeley
Quantum Nanoelectronics Laboratory ~\cite{qnl}, with eight
superconducting qubits connected in a line topology and with a native
gate set consisting of rotations and \cn gates \{$R_x(90),
  R_z(\theta), CNOT$\}. We then consider a chip with better connectivity, such as the ``bowtie'' employed in
  the IBM Q 5~\cite{ibmq5}, but a similar native gate set. Note that other superconducting qubit architectures support very similar  native gate sets, some with restricted topology~\cite{tqfrigetti}, others with richer connectivity~\cite{Google-2018}.  }

While topology is important for superconducting qubits, implementations using trapped ion~\cite{itq} qubits provide all-to-all connectivity through M\o lmer-S\o rensen~\cite{msg}  gates.

\section{Synthesis Algorithm}
\label{sec:alg}
\newcommand{\ui}{U_{implemented} }
\newcommand{\uis}{$U_{implemented}$ }
\newcommand{\ut}{U_{target} }
\newcommand{\uts}{$U_{target}$}
Our goal is to design an algorithm that addresses currently  perceived
shortcomings of synthesis  
and that  can be  easily extended to new hardware in order  to enable design space exploration in
quantum programming. To be useful during the NISQ device era we use \cn count as our primary optimality criteria.  The synthesis algorithm described in the rest of this section combines a generalized space of parameterized circuits with an approximate A* search~\cite{astar}.

\begin{figure*}
  \begin{tabular}{c}
\begin{minipage}{\textwidth}
  \centerline{\includegraphics[width=3in,height=1in,keepaspectratio]{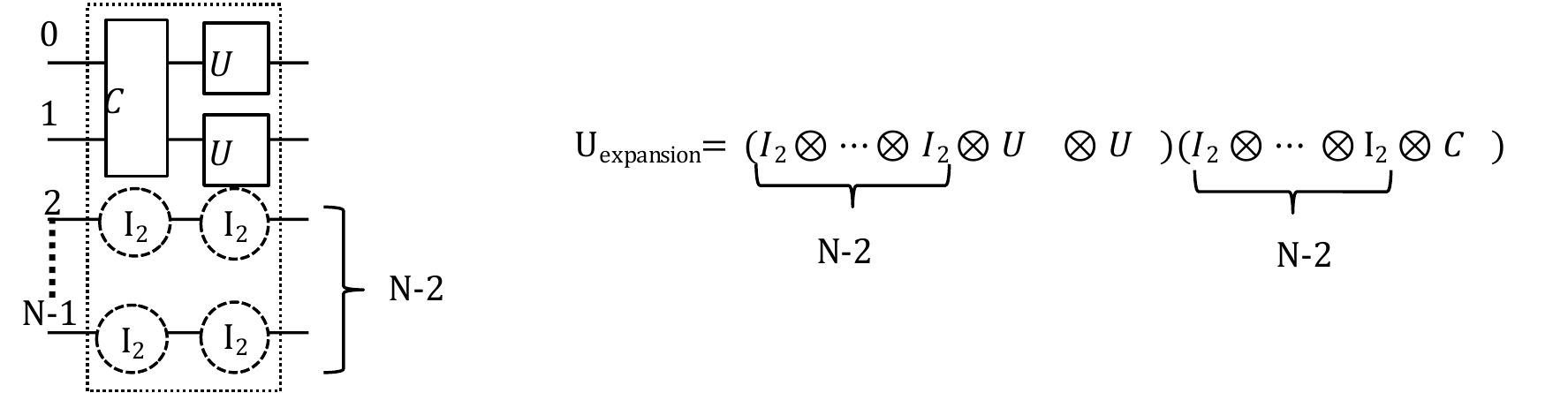}}
  \centerline{(A)}
\end{minipage} \\

    \begin{minipage}{\textwidth}
  \centerline{\includegraphics[width=5in,height=3in]{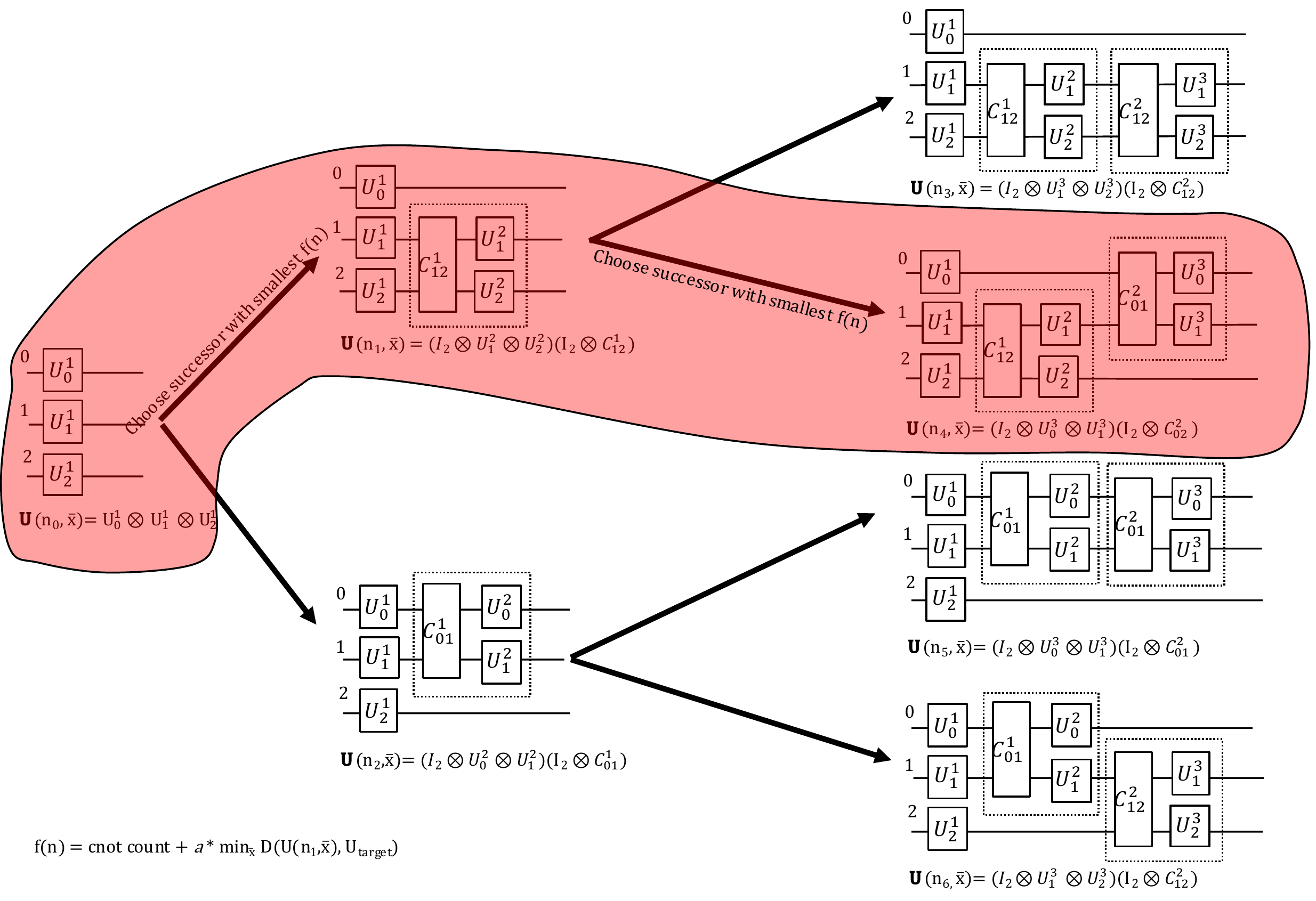}}
 \centerline{ (B)}
  \caption{\label{fig:usearch} \footnotesize \it  (A) Basic circuit block used for
    expanding the solution. We generate all alternatives where this
    structure is placed on linked qubit pairs. Each step adds six
    additional parameters to the optimization problem.  (B) Example evolution
    of the search algorithm for a three qubit circuit. We start by placing a layer of single
    qubit gates, then generate the next two possible solutions. Each
    is evaluated and in this case the upper circuit is closer to the
    target unitary, leading to a smaller heuristic value. Since this circuit This circuit is then expanded with its possible
    two successors. These are again instantiated by the optimizer. The
    second circuit from the top has an acceptable distance and is
    reported as the solution. The path in red  shows the evolution of
    the solution. The solutions enclosed by the dotted line have been
    evaluated during the search.}
 \end{minipage} \\
 \end{tabular}
\end{figure*}

Intuitively, search based synthesis methods rely on the following approach. They start by ``enumerating'' the space of possible solutions. The construction of this space guarantees that if a solution exists, it will be contained in the enumeration. We rely on the same strategy. 
Then they start walking this space looking  for solutions. Previous work uses ``randomized'' walk through genetic algorithms or Monte-Carlo methods. In contrast, we use a more regimented approach where we formulate the problem as a graph search and deploy established algorithms with good properties. In our case this is the A* algorithm. 
An example of the evolution of a search on a three qubit circuit is depicted in Figure~\ref{fig:usearch}.

\subsection{Formulation of Synthesis as a Tree Search Problem}

We first formulate the problem of synthesis as a graph search problem. We do this by constructing a tree of circuit structures.

The root node of our tree consists of $U3$ gates on every qubit line. For each node in the tree, there is one child for each possible \cn position. For each \cn position, we can construct the child by adding a \cn in that position, and then adding two $U3$ gates on the qubit lines affected by the new CNOT.

For any circuit that can be constructed with a finite number of \cn and $U3$ gates, our tree contains a node that can represent it. We will now provide constructive proof.

As a base case, the empty circuit, which contains 0 gates, implements the identity matrix, which can be represented by the root node with zero for all of its parameters, which also implements the identity. Now, assume that we can represent all circuits with up to $i$ gates. Given a circuit of length $i+1$, we can take the first $i$ gates, and find the node in our tree for it. For the last gate, if it is a \cn, we can represent it by choosing the child of the node for the first $i$ gates that appends a \cn in that position, and set the parameters of the two following $U3$ gates to 0. If the last gate is a $U3$, notice that the last gate on every qubit line in our circuit structure for any of our nodes is a $U3$ gate. The root node contains solely $U3$ gates, and any node further down the tree builds on the root node, so no qubit line is empty. The last gate on a qubit line is never a \cn because we add $U3$ gates immediately after every \cn. Therefore, the last $U3$ of the $i+1$ circuit is next to a $U3$ gate in the $i$ circuit, and we can combine these two $U3$ gates into a single $U3$ gate with different parameters, and we can use the same node in the tree. In any case, we have found a representation of the circuit of length $i+1$ in our tree.

The gate-set of $U3$ and \cn is universal for quantum computing, meaning that any unitary matrix can be represented by a circuit consisting of only those gates. Since our tree contains a representation of any such circuit, our tree can represent a circuit that implements any given unitary. Furthermore, since our tree is organized such that circuits with fewer \cn gates have a lower depth, if we find a lowest depth circuit that implements a given unitary, it will be a solution of lowest \cn count. We have now reduced the problem of finding a circuit for a given unitary with the lowest \cn count to a tree search problem, and then the numerical problem of finding values for the parameters. The first problem we can solve via A* search, and the second we can solve using numerical optimizer methods.

\subsection{The Synthesis Algorithm}
\newcommand{\sn}{$s(n)$}
\newcommand{\pn}{$p(n, \ut)$}
Our algorithm begins with a target unitary \uts, and a target gate-set. It also requires an acceptability threshold $\epsilon$, and a \cn count limit $\delta$.
The threshold provides the optimality metric for the solution.  The CNOT count limit ensures termination and it is selected as depth bounds provided by other competing~\cite{raban} methods: if we haven't found a solution there are better methods available and we stop. The following description refers to the pseudocode in Algorithms~\ref{fig:alg} and \ref{fig:help}. 

The algorithm relies heavily on a successor function \sn, which takes a node as input and returns a list of nodes, and an optimization function \pn, which takes a node and a unitary as input and returns a distance value. The function $H(d)$ is a heuristic function employed by A*, described in the next section.

The successor function, \sn, is defined based on the target gate-set and topology. Given a node $n$ as input, \sn generates a successor by appending to the circuit structure described by $n$.  It appends a \cn followed by two $U3$ gates. One successor is generated for each possible placement of the two-qubit gates allowed by the given topology.  The one-qubit gates are placed immediately after the CNOT, on the qubit lines that the \cn affects.  A list of all successors generated from $n$ this way is returned. Note that \cn and $U3$ can be replaced by different gates when using a different gate-set, as long as the gate-set remains universal and the single qubit gates are parameterizations of SU(2).

The optimization function, \pn, is used to find the closest matching circuit to a target unitary given a circuit structure. Given a node $n$ and a unitary \uts, let $U(n, \overline{x})$ represent the unitary implemented by the circuit structure represented by $n$ when using the vector $\overline{x}$ as parameters for the parameterized gates in the circuit structure. $D(U(n,\overline{x}),\ut)$ is used as an objective function, and is given to a numerical optimizer, which finds $d = {min}_{\overline{x}} D(U(n, \overline{x}), \ut)$. The function \pn returns $d$.

The algorithm begins by generating the root node, which describes a circuit structure with one $U3$ gate on each qubit line.  The distance value is found for the root node using \pn. These variables are initialized using the root node. The algorithm creates a priority queue that chooses the node $n$ that minimizes $f(n)$, and initializes the queue with the root node as the first entry. Now the algorithm enters a loop, in which it pops nodes from the queue.  If no node remains in the queue, the algorithm exits with no solution.  Otherwise, a node $n$ is successfully popped from the queue.  Its successors $n_1$-$n_k$ are generated using \sn. For each successor node $n_i$, the distance $d_i = p(n_i)$ is calculated: this is a source of parallelism. If $d_i < \epsilon$, the current circuit $n_i$ is deemed acceptable and is returned. Otherwise, if the \cn count of $n_i$ is within the limit $\delta$, the node $n_i$ is pushed onto the priority queue. If there is no acceptable solution with fewer than $\delta$ \cn gates, eventually all possible structures with fewer gates will be tried, the queue will empty, and no solution will be returned.

The node that is returned from the algorithm, $n_{final}$, represents a circuit structure that includes a circuit that implements \uts to a distance within $\epsilon$. To find the specific circuit, the same numerical optimizer can be used, but this time to find $\overline{x}_m = {arg\,min}_{\overline{x}} D(U(n, \overline{x}), \ut)$. In practice, it is not necessary to re-run the optimizer since optimizer functions generally return both the minimum value and the values of the parameters that minimize it. The pair of $n_{final}$ and $\overline{x}_m$ constitute a complete description of a quantum circuit, and can be directly converted to quantum assembly.

 \begin{figure*}
   \begin{minipage}{\textwidth}
     \begin{algorithm}[H]
 \caption{\label{fig:help} Helper Functions}
 \begin{algorithmic}[1]
   \Function{s}{$n$}
 \Return $\{n + \cn + U3 \otimes U3 \text{for all possible \cn positions}\}$
 \EndFunction
 \\
 \Function{p}{$n$, $U$}
 \Return ${min}_{\overline{x}} D(U(n, \overline{x}), U)$
 \EndFunction
 \\
 \Function{h}{$d$}\\
 \Return $d * a$ \Comment{$a$ is a constant determined via experiment. See section 3.3.1}
 \EndFunction
 \end{algorithmic}
 \end{algorithm}
 \begin{algorithm}[H]
 \caption{\label{fig:alg} Search Synthesis}
 \begin{algorithmic}[1]
 \Function{synthesize}{\uts, $\epsilon$, $\delta$}
 \State $n \gets \text{representation of $U3$ on each qubit}$
 \State \textbf{push } $n$ \textbf{ onto } $queue$ \textbf{ with priority } \Call{h}{$d_{best}$}$ + 0$
 \While{$queue$ is not empty}
 \State $n \gets \textbf{pop from } queue$
 \ForAll{$n_i \in $\Call{s}{$n$}}
 \State $d_i \gets$ \Call{p}{$n_i$, \uts}
 \If{$d_i < \epsilon$}
 \State \textbf{return }$n_i$
 \EndIf
 \If {\cn count of $n_i < \delta$}
 \State \textbf{push } $n_i$ \textbf{ onto } $queue$ \textbf{ with priority } \Call{h}{$d_i$}$ + $\cn count of $n_i$
 \EndIf
 \EndFor
 \EndWhile
 \EndFunction
 \end{algorithmic}
\end{algorithm}

\end{minipage} \\
\end{figure*}

\subsection{A* Search Strategy}
	The A* algorithm has been developed for graph traversals and it attempts to find a path between a start and target node. At each step, a partial solution is expanded using a successor function, and the successors are added to a priority queue.  Then a new partial solution is chosen from the queue that minimizes a cost function.  The first path from start to finish is the final solution.  Given a partial solution, the algorithm picks the next partial solution based on the cost of its already computed path and an estimate of the cost required to extend it all the way to the target. A* selects the successor node $n$ that minimizes
$f(n) = g(n) + h(n)$  where
\begin{itemize}
\item $f(n)$ is the estimated total cost of the path from start to finish
\item $g(n)$ is the cost of the path from the start to $n$
\item $h(n)$ is a heuristic function that estimates the cost of the cheapest path
from $n$ to the target
\end{itemize}
The algorithm terminates when it reaches the target node or if there are no paths eligible to be extended. The heuristic function is problem-specific and directly determines the time complexity of A*. If the heuristic function is {\it admissible}, meaning that it never overestimates the actual cost to get to the target, A* is guaranteed to return a least-cost path. A* can be run with an inadmissible heuristic to obtain sub-optimal solutions with a faster runtime than it would take to obtain guaranteed optimal solutions.

For synthesis, the selection of $g(n)$ is obvious as the CNOT count of the partial solution $n$. The challenge was to determine the heuristic function $h(n)$. After several attempts at deriving it from first principles we have opted for a data-driven approach described below. 

\subsubsection{Heuristic Function Tuning}
We first use breadth first search for synthesis. We ran breadth first search on each of our three-qubit benchmarks, and examined the details of the search along the final paths. At each partial solution along the path, we recorded the distance value at that step to the remaining number of \cn gates (calculated by subtracting the current number of \cn's at that step to the final value reach in that run of the program). We then fit the data, and found a best fit line with slope $a=9.3623$.

The fit gives us the heuristic function $h(n) = D(U(n,\overline{x}_{m}),U_{target}) * 9.3623$, or $h(n) = p(n, \ut) * 9.3623$. Although the fit was not very well correlated ($r^2 = 0.4102$), we found experimentally that the heuristic yielded excellent results. Running the same set of benchmarks with the A* heuristic, we found that the same quality solutions were found, but runtime was significantly faster. For example, brute force search for three qubit QFT takes one hour, while A* takes only seven minutes.

\subsection{Unitary  Distance Metric}
We use the following distance function based on the Hilbert-Schmidt inner product. If N is the dimension of the unitaries,

\begin{equation}
  D(U, U_{target}) = 1 -  \frac{\langle U, U_{target} \rangle_{HS}}{N} = 1 -   \frac{\Trace(U^{\dag} U_{target})}{N} \\
  \label{eq:4}
\end{equation}

The formula is based on the fact that  the inverse of a unitary matrix is its conjugate transpose. If the synthesis succeeds and $U$ is not distinguishable from $U_{target}$, the product $U^{\dag}U_{target} = I_N$, where $I$ is the identity matrix.
Furthermore, the maximum magnitude that the trace of a unitary matrix can have is its size N, which occurs at the identity (up to a phase).
The closer $U^{\dag} U_{target}$ is to identity, the closer  $\frac{\Trace(U^{\dag} U_{target})}{N}$ is to $N$, thus the closer our distance function is to $0$.

Note that variations of formulas using  Hilbert-Schmidt inner product have been previously used in synthesis algorithms~\cite{qaqc,ionsynth}, and ours has the following properties
\begin{itemize}
\item The distance is 0 when compilation is exact.
\item  It is easy to compute.
\item It has operational meaning.
\end{itemize}


\section{Experimental Setup} 
\label{sec:exp}

\parah{Software Implementation} We implemented our algorithm in {\tt python} 3.7.4, using {\tt numpy} 1.14.4 for performing matrix multiplication, and the COBYLA numerical optimizer provided with {\tt scipy} 1.2.0. We use {\tt multiprocessing.Pool} for parallelism.  Most of the tests ran on a  single node of the Cori supercomputer hosted at the National Energy Research Scientific Computing Center (NERSC), where nodes contain two Intel Xeon E5-2698 v3 ("Haswell")  processors at 2.3 GHz  (32 cores total).

The implementation returns a circuit when it finds a solution with a Hilbert-Schmidt distance value less than $10^{-10}$ from the target. This is enough that the resulting
circuit is not distinguishable from the original, as well as avoiding numerical errors within the software stack. 

\comment{ An ideal compilation would have a distance of zero, but there will be some imperfection in the results of COBYLA and floating point error from repeated matrix operations, so a threshold is chosen such that the distance value for a successful compilation will be less than it.  Empirically, trial circuits that had a heuristic value greater than $10^{-10}$ were always greater than $10^{-2}$, while trial circuits with a value less than $10^{-10}$ were typically on the order of $10^{-13}$, so this threshold should be sufficient to distinguish states that only had nonzero error due to imperfections in calculating the parameters from states that had \cn placements that did not correctly implement the circuit.  \red{also maybe worth mentioning comparison of Hilbert-Schmidt distance to random vector comparison, and that compiling with different threshold limits doesn't have a real effect until the thresholds are like 0.01}}

\parah{Benchmarks}  We concentrate on algorithms spanning a small number of qubits as we are interested in compiling for the Noisy Intermediate-Scale Quantum devices. Our goal is to demonstrate the value of synthesis to practitioners under several usage scenarios: 1) compiling unitaries; 2) gate set design exploration;  and 3) circuit optimization.

The benchmark suite is composed of ``traditional'' algorithms used by other evaluation studies~\cite{noisemap}, and it contains the Quantum Fourier Transform~\cite{qft} algorithm, HHL~\cite{hhl}, Variational Quantum Eigensolver~\cite{McClean2015} algorithm, together with important quantum kernels such as Toffoli gates. In addition to qubit based circuits we consider the  qutrit circuits described in~\cite{qtrit}.

\parah{Experimental Results} A summary of the results is presented in Table~\ref{tab:res}. 
The columns labeled
\cn show our implementation, annotated with the topology of the target chip.   Besides circuit depth, we present
the Hilbert-Schmidt distance of the solution and total  compilation time. 

\parah{Customizing for QPU Gate Set and Topology} We target directly the gate set native to the quantum processor. Our initial implementation was tailored for the QNL8QR-v5 processor which supports in hardware the  {$R_x(90), R_z{\theta}, CNOT$} gates and its qubits are connected in a line topology. We have also re-targeted the algorithm for the IBM Q 5 qubit chip, with a similar native gate set but a bow-tie/triangle \begin{tikzpicture}[scale=0.5]
    \basetriangle
    
  \end{tikzpicture} topology. 

\parah{Use Cases} To showcase the extensibility of the proposed approach we  consider synthesis of qutrit gates, a problem of interest to hardware and algorithm designers.
To showcase the interaction between synthesis and the rest of the software development stack (optimizing compilers and mappers) we examine using  synthesis during the circuit optimization phase.  In addition, we are interested determining the impact of specializing the synthesis algorithm for a different topology.
For this we report the length of the synthesized circuits after being compiled and optimized using QISKit. For example.  the ``CNOT+QISKit''  label describes the experiment where we compile our generated circuit with QISKit.

\parah{Comparison with State-of-the-Art} In Table~\ref{tab:res},  the column labeled UQ shows the number of \cn generated by the UniversalQ~\cite{uq}  compiler, a state-of-the-art synthesis tool that uses internally multiple linear algebra based decomposition methods, including Cosine-Sine. For UQ, we report the best result obtained by any decomposition method available.

\section{ Compiling Unitaries to Circuits}
\label{sec:use}

\newcommand{\linett}{
 
  \coordinate (Y) at (0,0);
  \coordinate (Z) at (0,1);
  \fill[white!10, draw=black] (Y) -- (Z);
 \filldraw (Y) circle (3pt); \filldraw (Z) circle (3pt);
}

In all cases illustrated in Table~\ref{tab:res} we were able to
synthesize circuits shorter than the theoretical upper bounds provided
by ~\cite{raban}.  Their \cn count upper bounds for Q=2 and Q=3 are 3
and 20 respectively. When comparing against the UniversalQ compiler,
we generate significantly shorter circuits, using on average
$2.4\times$ fewer CNOTs, and as high as $5.3\times$.

Comparisons against other techniques are harder, due to lack of availability of
software implementations (some not released, some described only in
algorithmic form) and differences in native gate sets. When
comparing against~\cite{ionsynth}, they report no compilations shorter
than eight two-qubit gates (M\o lmer-S\o rensen) for a sample of 3-qubit random
unitaries. Shortest circuit obtained by our tool has 3 CNOT gates. Amy
et al~\cite{MIM13} report eight  CNOT gates for Toffoli, while our
implementation finds a circuit with only six CNOTs. They also report
not being able to synthesize three qubit QFT using less than 10 CNOT gates.

At small qubit count, perhaps the most important comparison is against
the depth obtained by hand optimization.  From this perspective our
algorithm behaves  well. For example, the optimal \cn count for
Toffoli~\cite{toffoli6} is six, which our algorithm matches. When
mapping to a linear topology, implementations introduce extra SWAPs,
up to a total of 12 CNOT gates. Our linear topology Toffoli contains
only eight CNOTs. The Fredkin gate is usually implemented as Toffoli
sandwiched between two more CNOT gates.
Hand optimized Fredkin for linear topologies is available in Cirq~\cite{cirq} with
nine CNOTs, while our implementation uses only eight. On a well
connected IBM topology we synthesize a Fredkin using only seven
CNOTs: IBM QISKit will produce a circuit with eight CNOTs. 

The HHL implementation was obtained from the QNL8QR-v5 development
team. Mapped to a linear topology by hand, the circuit had seven CNOT
gates, while our implementation contains only three. 

For QFT, the best known implementations use two and six CNOTs, for two
and three qubit circuits respectively, assuming a well connected
topology.  In our case, we obtain circuits that are three and seven
long, respectively. After examining the resulting circuits, omitted
for brevity,  we
attribute the difference to limitations in the numerical optimizer
(COBYLA in this case). In the optimal circuit, there are places where
there are no single qubit gates between CNOT gates. It seems that all
numerical optimizers we have experimented with have trouble zeroing
these gates, thus leading to a slightly longer circuit. Note that we
do obtain good results for QFT on a line topology: best circuits have
nine CNOTs, while ours has eight.

 \comment{Similarly,~\cite{noisemap} use a Fredkin implementation that
  is eight deep, while we obtain a depth of nine. These two cases
  expose a very subtle limitation of all numerical optimization based synthesis
  techniques currently published.

  As described in Section~\ref{sec:related}, modulo the expansion
  strategy, all  numerical optimization techniques alternate layers of CNOT and arbitrary
  single qubit gates in an attempt to reduce the dimensions of the
  search space. On the other hand, when examining the optimal
  implementations of Toffoli and Fredkin, the circuits contain a
  number of single qubit gates  lower than in any circuit synthesized by
  numerical optimization approaches. For Fredkin, the implementation used
  in~\cite{noisemap} uses 11 single qubit gates, while our algorithm
  will insert $3+8 \times 2 = 19$ gates. A similar situation occurs
  for Toffoli. In the ideal case, one would expect the numerical
  optimizer to be able to synthesize identity gates wherever
  necessary, but this does not seem to be the case. We are working to
  empirically validate this conjecture. If true, this insight
  indicates that synthesis can be improved by either more informed
  expansion strategies or by customized numerical optimization algorithms.}

\subsection{Impact of Topology}
Embedding the circuit topology within the synthesis algorithm matters,
perhaps even more than developing an optimal algorithm for well
connected topologies.

The first observation is that existing algorithms report large
($4\times$) proportionality constants when specializing for a
restricted topology. In our case we observe only modest increases, up
to 15\% for the workload and for only five of the tested circuits.
In some cases, we obtain circuits shorter than previously known.
This indicates that we can handle well restricted topologies.

Even more important is the empirical observation that the rest of the
compilation toolkit (circuit optimizers + mappers) can only  increase
(never decrease)  
the depth of our synthesized circuits. This is illustrated in
Table~\ref{tab:res} by the columns with the label ``QISKit''.
In the first experiment, we take the circuits synthesized for a linear
topology and compile them with QISKit for the better connected bowtie
topology. We enable the highest level of optimization available. The
circuits optimized and mapped by QISKit have the same length as the
input circuits. In the second experiment, data presented in the Table,
we compile the circuits synthesized for the bowtie with QISKit configured for
a linear topology. In this case we observe a 53\% average increase in CNOT
count, with values as high as $4\times$.

To us,
this indicates that if the goal for NISQ devices is obtaining
optimally short circuits, techniques like our are more likely to
deliver consistently than traditional optimizers and mappers.

\subsection{Synthesis and Circuit Optimization}

The three qubit circuit ``EntangledX'' provides an illustration of the
benefits of synthesis embedded in the circuit optimization
workflow. The gate is a building block for a  VQE
implementation using the [[4,2,2]] error detection code~\cite{422c}
and it is parameterized by a rotation angle. The authors run the
circuit sampling the parameter for robust behavior, the sampling is
directed by the results of the previous run.

The (painstakingly) hand optimized and most generic version contains
four CNOT gates, which we  match for most values of the rotation angle. However, for some angles, we were able to achieve circuits with only two or three \cn gates.

\comment{
The need for aggressive optimizations is further motivated by the emergence of
circuit generators. At the application or domain science layer, packages such as
OpenFermion~\cite{openfermion} produce circuits by translating from
physics or chemistry  first
principles. Due to the complexity of the
problem, the generated circuits are often long and suboptimal. At the
compiler level, it is worth exploring approaches for ``global''
circuit optimization, where circuits (or partitions thereof) are
lifted to unitaries and re-synthesized.

In Table~\ref{tab:res} the entry labeled ``VQE ethylene'' corresponds
to a four qubit circuit generated by OpenFermion to simulate  the ethylene
molecule. The circuit contains 64 \cn gates. We
lifted the circuit to the unitary matrix and attempted to
re-synthesize it. As illustrated, we have aborted the compilation at a
depth of 27 due to its long duration. As the bottleneck is the
numerical optimizer performance at large parameter count, circuit
partitioning techniques are needed when increasing the qubit count.}

\comment{
\begin{figure}
\tiny
\begin{tabular}{|c|c|c|c|c|}
\hline
 & Q & CNOT & UQ & TIME \\
\hline
QFT & 2 & 3 & 3 & 3.43 \\
\hline
QFT & 3 & 8 & 15 & 610.16 \\
Fredkin & 3 & 8 & 14 & 493.67 \\
Toffoli & 3 & 8 & 9 & 714.36 \\
Peres & 3 & 7 & 19 & 331.45 \\
Or & 3 & 8 & 10 & 492.77 \\
Miro & 3 & 4 & 9 & 27.35 \\
HHL & 3 & 3 & 16 & 12.73 \\
\hline
QFT & 4 & - & 89 & - \\
Full Adder & 4 & - & 94 & - \\
VQE Ethylene & 4 & - & 13 & -\\
\hline
\end{tabular}
\end{figure}
\begin{figure}
\tiny
\begin{tabular}{|c|c|c|c|c|}
\hline
ALG & Q & CNOT & UQ* & TIME (s) \\
\hline
QFT & 3 & 7 & 15 & 341.58 \\
Fredkin & 3 & 7 & 14 & 849.09 \\
Toffoli & 3 & 6 & 9 & 1015.32 \\
Peres & 3 & 6 & 19 & 285.86 \\
Or & 3 & 6 & 10 & 340.58 \\
Miro & 3 & 4 & 9 & 60.52 \\
HHL & 3 & 3 & 16 & 12.16 \\
\hline
\end{tabular}
\end{figure}
}

 \begin{figure*}
   \tiny
  \begin{tabular}{|c|c||c|c|c||c|c|c||c|c||c|c|c|}
    \hline& &\multicolumn{3}{c||}{\cn count} &
                                               \multicolumn{3}{c||}{Mapped
                                               by QISKit on linear topology}
    & \multicolumn{2}{c||}{Unitary distance} &
                                             \multicolumn{3}{c|}{Compile
                                             time (s)} \\
    
    \hline
    ALG & Qubits   & CNOT  \begin{tikzpicture}[scale=0.3]
    \linett
    \end{tikzpicture}  & CNOT \begin{tikzpicture}[scale=0.3]
    \basetriangle
    
  \end{tikzpicture}  &
                       UQ &
                         $CNOT $ \begin{tikzpicture}[scale=0.3]
                           \linett \end{tikzpicture} + QISKit &
                        $CNOT$ \begin{tikzpicture}[scale=0.3]
                          \basetriangle \end{tikzpicture} + QISKit &
                                                                     UQ+QISKit &
                                                                                 $|| CNOT ||$ \begin{tikzpicture}[scale=0.3]
                                                                                   \linett \end{tikzpicture} &
                          $||CNOT||$  \begin{tikzpicture}[scale=0.3]
                            \basetriangle \end{tikzpicture} &
                                                                       T($CNOT$) \begin{tikzpicture}[scale=0.3]
                           \linett \end{tikzpicture} 
                                                                       
                                                                   &
                                                                     T($CNOT$) \begin{tikzpicture}[scale=0.3]
                          \basetriangle \end{tikzpicture} & T(UQ) \\
    \hline
    \hline
    QFT &2 &3& 3& 3 & 3 & 3 & 3 & $2.08*10^{-15}$ & $3.86*10^{-15}$ &  3 & 3 & <1\\
    \hline
    \hline
    QFT &3&8&7&15&8& 13&  27& $1.56*10^{-14}$ & $8.66*10^{-15}$ &  610&341 & <1\\
    \hline
    Fredkin & 3 & 8 &  7& 9 & 8& 16& 26& $2.22*10^{-15}$ & $4.69*10^{-15}$ & 493&849 & <1\\
    \hline
Toffoli & 3 & 8 &  6 &9  &8& 12& 21& $1.10*10^{-14}$ & $1.88*10^{-14}$ & 714& 1015& <1\\
    \hline
 Peres&3&7&6&19  & 7& 9& 47 & $1.01*10^{-14}$ & $8.25*10^{-15} $ & 331& 285& <1\\
   
    \hline
HHL &3 &3&3 &16   & 3& 3&21 & $2.46*10^{-14}$ & $1.44*10^{-16}$ & 12&12 & <1\\
    \hline
    Or & 3 & 8&  6 & 10  & 8& 9& 19& $4.01*10^{-14}$ & $8.65*10^{-15}$ & 492&340 & <1\\
    \hline
    EntangledX & 3 & 4&  4 & 9  &4 & 16&21 & $7.77*10^{-17}$ & $2.11*10^{-16}$ & 27&60 & <1\\
    \hline
    \hline
    QFT & 4 & 14 & & 89 & & & DNR& $1.41 * 10^{-12}$ & & 410250 & & <1\\
    \hline
  \end{tabular}
  \caption{\label{tab:res}\it \footnotesize  Summary of synthesis results for
    several algorithms and unitaries. Q = number of qubits. The
    topology used during synthesis is denoted in the
    caption. Theoretical \cn count upper bounds for Q=2 and Q=3 are 3
    and 20 respectively. 
 }
\end{figure*}

\subsection{Retargeting to Qutrits}

Qutrits extend qubits to systems with three logical values 0, 1 and
2. They are represented by unitaries from SU(3) and extend from binary
to ternary logic to 
explore a space with $3^n$ dimensions. There exist several~\cite{qtsynth,qtparam}
decompositions and parameterizations, all using eight independent
parameters. Gates to implement qutrit operations  have been explored only
recently~\cite{qtrit} for qubit based systems, mostly motivated by
the need~\cite{bhscramble}  for
modeling  physical phenomena. 

For our study, we implement a CSUM two-qutrit gate, which adds the
value of the first qutrit to the second qutrit $CSUM(\ket{11}) = \ket{12}$.  Our synthesis matches
the hand optimized implementation  by~\cite{qtrit}.  For brevity, we
omit detailed results.

\subsection{Acceptability Threshold Tuning}
Our algorithm terminates upon finding a circuit with a distance value
within an acceptability threshold $\epsilon$. Its value is determined
by two requirements: 1) the implementation should be able to meet it
in terms of numerical accuracy; and 2) the resulting unitary should be
indistinguishable from the original.

For the first criteria we tried synthesizing four of our benchmarks
threshold limits at powers of ten ranging from $0.1$ to $10^{-12}$. We
found that with only two exceptions, threshold limits in the range
$0.01 \leq \epsilon \leq 10^{-12}$ resulted in final solutions with
distance on the order of $10^{-12} - 10^{-14}$. The two exceptions
were both 3-qubit QFT solutions, one with a solution on the order of
$10^{-10}$ and one with $10^{-8}$. We concluded that a threshold of
$10^{-10}$ will ensure we have the best quality answer our numerical
optimizers will be able to give us.
	
To ensure this threshold is sufficient for real world applications, we
ran another experiment to relate matrix distance to the KL divergence
of probability distributions. We generated random unitaries that are
close to the identity, multiplied these by fully random unitaries. For
each pair of fully random unitary and product of random unitary and
near-identity random unitary, we recorded the matrix distance and the
KL divergence between the final probability distributions after
measuring the result of applying the two unitaries to the same
randomly generated state vector, recording the worst case KL
divergence after trying 1000 random state vectors. The results showed
a clear correlation between KL divergence and Hilbert-Schmidt distance,
with the acceptability threshold of $10^{-10}$ yielding a maximum KL
divergence of $2.56*10^{-9}$. Even for a looser threshold of
$10^{-8}$, the maximum KL divergence was $5.20*10^{-8}$, so the threshold
might even be loosened in practice.

\subsection{Solution Quality}

The Hilbert-Schmidt distance between our solution and the original
unitary is presented in Table~\ref{tab:res}. The values range from
$10^{-14}$ to $10^{-17}$. We tested the resulting circuits on 1,000
random input state vectors: the results are indistinguishable from
the original circuit.

We only report the total number of CNOT gates in the generated
circuit. The upper bound on the total number of gates in a $Q$ qubits
circuit  is given by $Q+5*CNOT$. This includes single qubit gates and
is based on the fact that each $U_3$ gate is expanded into at most
four~\footnote{Normal decomposition ZXZXZ suggests five, but we use
  commutativity laws to move gates though CNOTs.}
single qubit rotations. 

\subsection{Running Time}

The running time of our algorithm is presented in
Table~\ref{tab:res}.
In the current implementation the algorithm performance is determined
mainly by the performance of the numerical optimizer. We have
experimented with several Python interoperable implementations: CMA-ES~\cite{cmaes},
COBYLA and BOBYQA. We have selected COBYLA as the default optimizer.
Given  a circuit with $Q$ qubits and depth $d$, the total number of
parameters is $Num\_Params = 3*Q + 6*d$. For reference, typical durations for depth
six circuits are $\approx 130s$, $\approx
210s$ at depth eight and $\approx 620s$ at depth 14. 

Our implementation is otherwise very well optimized. Some techniques are Python specific, some are generally applicable. We used our
  own object-based representation of quantum gates which allowed us a
  simple and memory-efficient way of implementing the successor
  function. These objects create and multiply together {\tt numpy} matrices and  we have thoroughly optimized the code to minimize object copies. We also  added a circuit-optimization step which rearranges our circuit component graph to perform matrix products with the minimum number of operations. The gate parametrization is minimized by  replacing the parameterized single qubit gate after the control line of a \cn gate with a simpler parameterization with only two parameters (because a parameterized Z gate can commute through the control line of a \cn and can be absorbed by the parameterized gate on the other side). The vast majority of our runtime is spent in creating matrices from circuit component graphs within the objective function calls of the optimizer, so we have focused our optimization efforts on tuning the optimizer to make fewer objective function calls and improving our matrix generation to be more efficient. We also used beam searching, popping multiple nodes off the top of the queue at a time, in order to take better advantage of parallelism. Beam searching lets us evaluate nodes that we would have to backtrack to in parallel rather than sequentially. In the case of approximate A*, it can lead to a different solution, but it will only find one at least as good (in terms of minimizing \cn count) as it would have found otherwise.

Note that some of the runtime overhead cannot be avoided in a Python code base, but disappears when re-implementing in a performance oriented language such as C/C++.  We have chosen Python for the easy interoperability with all available~\cite{qiskit,cirq,pyquil}  quantum compilation infrastructures.

\section{Discussion}
\label{sec:disc}

Overall, we believe our results are very encouraging and show the
general applicability of quantum circuit synthesis techniques during
the NISQ decade(s). Looking back, the field has progressed
steadily. Solovay and Kitaev open the field by showing that a solution
exists when using any universal gate set. Later efforts show that
solutions exist when restricting the gate sets to ``almost
native''. The emphasis then moved on to improving quality (depth) of
the solution, and the field has steadily progressed from computing 
huge~\footnote{Our own Solovay Kitaev implementation synthesized a two
  qubit gate with depth 10,000. Optimal depth is at most at 3 CNOT.}
to computing decent solutions.  

We have shown concrete results where we match the
shortest known depth for several algorithms, we have shown results
where we reduce depth for constrained topologies (line) and we have
shown the retargetability of the implementation to new gate
sets. Equally important, we have shown empirical evidence that traditional optimization 
techniques (peephole optimizers and mappers) are unlikely to match the
quality of the circuits generated by synthesis. We believe that the
results alleviate some of the doubts faced by synthesis approaches:
generated circuits are too deep and there is no topology awareness.  

\comment{
We attain extensibility  by a combination of design
choices.  First, we target directly the hardware native gate set. This is easily retargetable, as illustrated by the set of
experiments using $\sqrt{CNOT}$ gates.
Topology awareness is attained by using
directly the information about chip connectivity  when devising the
part of the search strategy that attempts to expand on the frontier of
partial solutions. This is also easily retargetable as illustrated by
the experiments on the line and bowtie topologies. }

Due to its potential,  we believe a roadmap for synthesis targeting
NISQ devices
is worth developing. Our study illustrates some of the solutions, as
well as the associated open problems. For practical purposes, quality
of the solution is important (short depth), followed by
scalability. Given that we have shown optimality and topology
awareness,  for the near
future, scalability at small qubit scale is worth exploring as it will
lead to establishing robust building blocks when considering
scalability with qubits.

\parah{Synthesis for early NISQ (small) circuits} There are several orthogonal directions to pursue to improve speed:
\begin{itemize}
  
  \item {\it Better numerical optimizers.} The judicious choice of the numerical optimizer is probably the most
important factor. ~\cite{Rios2013} provide a very useful overview of
derivative-free methods. Based on their recommendations, the first
step is to select the best known derivative-free methods such as
TOMLAB/MULTIMIN, TOMLAB/GLCCLUSTER, MCS or TOMLAB/LGO. The second
step is to employ meta-optimization techniques that combine different
approaches. Note that in our case the choice was limited due
to lack of availability of open source Python or C based
implementations. It is  also worth considering building
ad-hoc optimizers for synthesis based on tensor networks and gradient descent. These have
the advantage of high GPU performance. 

\item {\it Better parallelization of the search algorithm.} There are
  two levels of parallelism within numerical  optimization based algorithms. At
  the inner level, the
  first challenge is that the numerical 
  optimizer itself needs to have a good parallel implementation.This
  does not seem to be the case with the publicly available
  implementations which  exploit it only in small matrix BLAS function
  calls. There is  an outer level of embarrassing parallelism across
  optimizer invocations, given by the evaluation of the partial
  solutions at a given search step. This is proportional with the number of
  qubits in the algorithm.  Since in our case the optimizer performs
  best single threaded, shared memory parallelism is
  sufficient. Implementations will eventually need to move to distributed memory
  parallelism, given the availability of parallel numerical optimizers.

\comment{
  The current
  approaches do not offer any level of coordination between the
  parallel evaluations of solutions. One possible direction is to
  develop more coordinated algorithms for parallel searches.
  All existing search approaches assume the cost per step is
independent of depth in the solution tree. Another direction worth pursuing seems to be developing algorithms
  that guide the search using a heuristic that combines solution
  quality with expansion cost.   One idea is to develop hybrid techniques that combine
  breadth-first with depth-first traversals that account for the
  number of parameters, as inspired by~\cite{Beamer2012} in the graph
  algorithms realm.  \red{A lot of this I think could be improved.  I'll write my suggestion for this blurb later.  Also a comment earlier said to find this in section 7, but it ended up being in section 8.}

  \item Methods to bound the number of parameters.  In our opinion these are the most
    promising in terms of improving speed to solution. Our initial
    attempts to recognize and ``memoize'' blocks of parameters that
    are slowly changing during synthesis failed, but there are plenty
    of unexplored venues in this direction.  Another possible
    direction is to consider hybrid approaches that combine matrix
    decomposition methods such as Cosine-Sine with numerical
    optimization methods on sub-terms. This is one of our areas of
    future work.
 \end{itemize}

 When considering solution optimality as measured by the depth of
 generated circuit, one observation is that methods based on numerical
 optimization do not consider ancilla qubits. Another area of
 improvement is development of such methods. In our opinion, the
 easiest way is combining decomposition methods (Cosine-Sine) which
 already can handle ancillas with numerical optimization. \red{If we want to include ancillas, the approach would be to modify our distance function to ignore one or more qubits.  Our research shows that Cosine-Sine is pretty bad and probably should not be something that we use if we can avoid it.}
 } 
  \parah{Synthesis for late NISQ (large) circuits} For circuits with
  tens of qubits memory and computational requirements for synthesis
  may be prohibitive, as unitaries scale exponentially with $2^q$. Given
  an already existing circuit, a straightforward way to incorporate
  synthesis is to partition it in manageable size blocks, optimize
  these individually and recombine. For algorithm discovery, synthesis
  will have to be incorporated into generative models for domain
  science. For example, frameworks such as OpenFermion can already
  generate arbitrary size circuits.  We have already started exploring
  these directions using the current algorithm.

\section{Related Work}
\label{sec:related}

A fundamental result, which
spurred the apparition of quantum circuit synthesis is provided by the
Solovay Kitaev (SK) theorem.The theorem
relates circuit depth to the quality of the approximation and its
proof is by construction ~\cite{DawsonNielson05,Nagy16,ola15}. Different approaches~\cite{DawsonNielson05,ZXZ16,BocharovPRL12,MIM13,Qcompile16,ctmq,23gates,householderQ,CSD04,amy16,seroussi80} to synthesis have been
introduced since, with the goal of generating shorter depth circuits. 
These can be coarsely classified based on several
criteria: 1) target gate set; 2) algorithmic approach; and 3) solution distinguishability.

\parah{Target Gate Set} The SK algorithm is quite general in the sense that it is applicable to any 
universal gate set. Synthesis can be improved in terms of both speed
and  optimality by specializing the gate set. Examples include
synthesis of z-rotation unitaries with
Clifford+V approximation~\cite{Ross15} or Clifford+T gates~\cite{KMM16}. When ancillary qubits are allowed, one can synthesize
a single qubit unitaries  with the Clifford+T gate set~\cite{KMM16,KSV02,Paetznick2014}. 
While these efforts propelled the field of synthesis, they are not 
used on NISQ devices, which offer a different gate set
($R_x, R_z,CNOT$ and M\o lmer-S\o rensen all-to-all).
Several~\cite{raban,synthcsd,ionsynth}  other
algorithms, discussed below have since emerged.

\comment{ For example, the z-rotation unitaries can be
synthesized with Clifford+V approximation~\cite{Ross15} or  with Clifford+T gates~\cite{KMM16}. The set of Pauli, Hadamard, Phase, CNOT 
gates form what is known as the Clifford group gates. When augmented with the T gate defined
as 
\[
T = \left(
\begin{array}{cc}
1 & 0 \\
0 & \zeta_8 
\end{array}
\right), \ \ \mbox{where} \ \ \zeta_8 = e^{i\pi/4},
\]
the gate set is universal.
t lead to better complexity $\mathcal{O}(\log^{1.75}(1/\epsilon))$ compared 
to the SK Algorithm.  \comment{$\mathcal{O}(\log(1/\epsilon))$ T-count scaling.}
}  

\parah{Algorithmic Approaches} The earlier attempts inspired by
Solovay Kitaev use a recursive (or divide-an-conquer) 
formulation, sometimes supplemented with search heuristics at the
bottom. More recent search based approaches are illustrated by the
Meet-in-the-Middle~\cite{MIM13}  algorithm.

Several  approaches use techniques from linear algebra for
unitary/tensor decomposition.  \cite{23gates} use QR matrix factorization via Given's rotation and Householder transformation
~\cite{householderQ}, but there are open questions as to the suitability
for hardware implementation because these algorithms are expressed 
in terms of row and column updates of a matrix rather than in terms of qubits.

The state-of-the-art upper bounds on circuit depth are provided by
techniques~\cite{synthcsd,raban} that use Cosine-Sine
decomposition. The Cosine-Sine decomposition was first
used by~\cite{tucci}  for compilation purposes. In practice,
commercial compilers ubiquitously deploy only 
KAK~\cite{KAK} decompositions for two qubit unitaries.

The basic formulation of these techniques is topology
independent. Specializing for topology increases the upper bound by
rather large constants, ~\cite{synthcsd} mention a factor of nine,
improved by~\cite{raban} to $4\times$. 
The published approaches are hard to extend to different qubit gate
sets and it remains to be seen if they can handle\footnote{~\cite{qtsynth} describes a method
using Givens rotations and Householder decomposition.} qutrits.  Furthermore, it seems that the numerical
techniques~\cite{computecsd} required for CSD still require
refinements as they cannot handle numerically challenging cases.

\comment{
There are not many studies published about synthesis of qutrit based
circuits and qutrit gate sets.~\cite{qtsynth} describes a method
using Givens rotations and Householder decomposition. As techniques
for qubit based systems using a similar approach have been
proposed~\cite{23gates}, they may allow an easier combination of
qutrit and qubit based synthesis. }

Several techniques use numerical optimization, much as we did. They
describe the gates in their variational/continuous representation and
use optimizers and search to find a gate decomposition and
instantiation.
The work closest to ours is by~\cite{ionsynth} which  use
numerical optimization and brute force search to synthesize circuits
for a processor using trapped ion qubits. Their main advantage is the
existence of all-to-all M\o lmer-S\o rensen gates, which allow a topology
independent approach. The main difference between our work and theirs
is that they use randomization and genetic algorithms to search the
solution space, while we show a more regimented way.
When Martinez et al. describe their results,
they claim that M\o lmer-S\o rensen counts are directly comparable to CNOT
counts. By this metric, we seem to generate comparable or shorter
circuits than theirs.  It is not clear how their approach behaves when
topology constraints are present. The direct comparison is further limited due to
the fact that they consider only randomly generated unitaries, rather
than algorithms or well understood gates such as Toffoli or Fredkin.

Another topology independent numerical optimization technique is
presented by~\cite{qaqc}. In this case, the main contribution is to
use a quantum annealer to do searches over sequences of increasing
gate depth. They report results only for two qubit circuits.

All existing studies focus on the quality of the solution, rather than
synthesis speed. They also report results for low qubit concurrency:
Khatri et al.~\cite{qaqc} for two qubit systems, Martinez et al.~\cite{ionsynth} for systems up to
four qubits.

\parah{Solution Distinguishability} 
Synthesis algorithms are classified as exact or approximate based on
distinguishability.  This is a subtle classification criteria, as most
algorithms can be viewed as either.  For example,~\cite{MIM13}
proposed a divide-and-conquer algorithm called Meet-in-the-Middle
(MIM). Designed for exact circuit synthesis, the algorithm
may also be used to construct an $\epsilon$-approximate circuit. The results seem to indicate that the algorithm failed
to synthesize a three qubit QFT circuit. 

Furthermore, on NISQ devices, the target gate set of the algorithm
(e.g. T gate) may
be itself implemented as an approximation when using native gates.

We classify our approach as approximate since we accept solutions at a small distance from the original
unitary. In a sense, when algorithms move from design to
implementation, all algorithms are approximate due to numerical
floating point errors.

\comment{
It allows one to search for
circuits of depth $l$ by only generating circuits of depth at most $\lceil l/2 \rceil$ at the complexity of $\mathcal{O}(|\mathcal{V}_{n,\mathcal{G}}|^{\lceil l/2\rceil}\log |\mathcal{V}_{n,\mathcal{G}}|^{\lceil l/2 \rceil})$, 
where $\mathcal{V}_{n,\mathcal{G}}$ denotes the set of unitaries for depth-one 
$n$-qubit circuit. The MIM algorithm is flexible and allows weights to be
added to the gate set to account for the possibility that some gates,
such as those that do not belong to the Clifford group, may be more expensive
to implement. It also allows ancillas to be used in the synthesis.  The 
algorithm uses a number of heuristics to prune the search tree.  Although
it was originally designed for exact circuit synthesis, the algorithm
may also be used to construct an $\epsilon$-approximate circuit.
}

\section{Conclusion}
\label{sec:conc}

In this work we have shown methods to compile arbitrary quantum
unitaries into a sequence of gates native to several superconducting
qubit based architectures. The algorithm we develop is topology aware
and it is easily re-targeted to new gates sets or topologies. Results
indicate that we can match, or even improve on when topology is restricted,
the shortest depth
circuit implementation published  for  several widely used algorithms
and gates. We also show empirical evidence which supports an important
conjecture: the benefits of incorporating topology directly into
synthesis cannot be replicated if relying on all-to-all synthesis and
traditional (peephole base) optimizing quantum compilers or mappers.  

The method is slow but it does produce  good results in
practice. For the early NISQ era, which is likely to be characterized
by hero experiments, the overhead seems acceptable. Even when
superseded by faster algorithms, we believe our results provide a good
quality  measure threshold for these implementations.

Looking forward,
better numerical optimizers are required for enhancing the
palatability of quantum circuit synthesis. These will alleviate some of the
need for developing better search algorithms.



\end{itemize}
\end{document}